\documentclass[conference]{IEEEtran}

\IEEEoverridecommandlockouts
\usepackage{booktabs}
\usepackage{soul}
\usepackage{cite}
\usepackage{amsmath,amssymb,amsfonts}
\usepackage{algorithmic}
\usepackage{graphicx}
\usepackage{textcomp}
\usepackage{xcolor}
\usepackage{float}
\usepackage{subcaption} % For subfigures
\usepackage{soul,color}

\newcommand{\eqlabel}[1]{\label{eq:#1}}
\renewcommand{\eqref}[1]{~(\ref{eq:#1})}

\def\BibTeX{{\rm B\kern-.05em{\sc i\kern-.025em b}\kern-.08em
		T\kern-.1667em\lower.7ex\hbox{E}\kern-.125emX}}
\graphicspath{{figures/}}
\begin{document}

\title{Event-Driven Simulation of Power Electronics Rich Grid Models
	\thanks{This manuscript has been authored by UT-Battelle, LLC, under contract DE-AC05-00OR22725 with the US Department of Energy (DOE). The US government retains and the publisher, by accepting the article for publication, acknowledges that the US government retains a nonexclusive, paid-up, irrevocable, worldwide license to publish or reproduce the published form of this manuscript, or allow others to do so, for US government purposes. DOE will provide public access to these results of federally sponsored research in accordance with the DOE Public Access Plan (http://energy.gov/downloads/doe-public-access-plan).}}

\author{\IEEEauthorblockN{Ajay Pratap Yadav}
\IEEEauthorblockA{\textit{Computational Sciences} \\
	\textit{\& Engineering Division} \\
	\textit{Oak Ridge National Laboratory} \\
	Oak Ridge, TN, USA \\
	yadavap@ornl.gov}
\and
\IEEEauthorblockN{James J. Nutaro}
\IEEEauthorblockA{\textit{Computational Sciences} \\
	\textit{\& Engineering Division} \\
	\textit{Oak Ridge National Laboratory} \\
	Oak Ridge, TN, USA \\
	nutarojj@ornl.gov}}

\maketitle

\begin{abstract}
Power-electronics systems should be treated according to their natural mathematical structure---inherent switching and discontinuities with piecewise continuous states. Therefore, the simulator should be organized around events, switching topologies, and topology intervals, rather than only around a continuous-time solver that later corrects or smooths discontinuities. This paper presents a simple, event-driven EMT architecture using native C kernels with Python orchestration. With this method, we distill the essential elements of discrete-event simulation applied to power electronics problems and thereby point toward a broad research thrust wherein mature ideas from discrete-event simulation are adapted for use in simulating power electronics circuits.

\end{abstract}

\begin{IEEEkeywords}
Electromagnetic transients, power electronics, inverter-based resources, event-driven simulation, hybrid systems, EMT acceleration. 
\end{IEEEkeywords}

\section{Introduction}
\label{sec:introduction}
In the last several years there have been several research efforts demonstrating the advantages of event-driven simulation algorithms for power electronics models. Some of the most prominent efforts in this direction are the Discrete State Event-Driven Framework proposed by \cite{8086106,8675318} and the use of quantized state numerical integration methods as described by \cite{qss_review}. The effectiveness of these methods is both remarkable and a fundamental break from the time-driven methods that predominate at the present.

\begin{figure}[t]
	\centering
	\includegraphics[width=0.95\columnwidth]{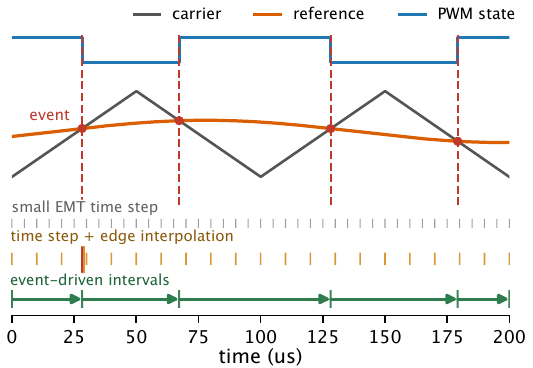}
	\caption{PWM events partition EMT dynamics into piecewise-continuous intervals. Fixed-step methods sample through the edge, interpolation corrects it, and event-driven simulation advances directly between topology events.}
	\label{fig:pwm_events}
\end{figure}

The radical shift in perspective inherent in these event-driven proposals, combined with the complexity of the algorithms involved, may hide their underlying similarities as discrete-event simulations \cite{zeigler}. Our aim is to distill the essence of discrete-event simulation when it is adapted for power electronics. Our approach resembles the split system method, in which a variable step numerical scheme is combined with a discrete-event simulation \cite{nutaro,10.1177/0037549711401000}. This framework makes explicit the two pieces of the simulation problem: locating events that cause discontinuous change in the circuit and solving the circuit equations between events.
Although simple, our event-driven algorithm shows significant speed up over state of the art approaches to time-driven simulation. This provides evidence that it is the event-driven approach to algorithm creation, more than its specific algorithmic features, that is the primary factor in speeding up the simulation. We also hope that a simplistic, but effective, algorithm will demystify the event-driven approach, making it more accessible to a power electronics focused research community.

\section{Event-Driven EMT Method}
\label{sec:method}
A converter-rich EMT model can be written as
\begin{equation}
	E \dot{x}=A_{\sigma}x+b_{\sigma}(u,t),
	\eqlabel{eq:emt_dae}
\end{equation}
where $x$ contains electrical and controller states, $E$ may be singular, and $A_{\sigma}$ and $b_{\sigma}$ change with switching topology $\sigma$ and input $u$. A minimal switched-converter example is
\begin{align}
	L\dot{i} &= s v_{\mathrm{dc}}-v_{\mathrm{grid}}-Ri, \eqlabel{eq:simple_l}\\
	C\dot{v}_{\mathrm{dc}} &= i_{\mathrm{pv}}-s i-\frac{v_{\mathrm{dc}}}{R},
	\eqlabel{eq:simple_c}
\end{align}
where $i$ is an inductor current, $v_{\mathrm{dc}}$ is the dc-link voltage, $s \in \{-1,0,1\}$ is the switching state, and $v_{\mathrm{grid}}$ and $i_{\mathrm{pv}}$ are the grid voltage and PV-side current. 
Switching events change $s$, or equivalently $\sigma$, and start new piecewise-continuous trajectory segments. Switch-level EMT therefore requires both accurate topology-event timing and stable propagation of the continuous dynamics between events.

\subsection{Event Calendar}
A time-driven EMT method advances by
\begin{equation}
	t_{k+1}=t_k+h.
\end{equation}
An event-driven method instead advances to the next physical or numerical event:
\begin{equation}
	t_{k+1}=\min(t_{\mathrm{gate}},t_{\mathrm{diode}},
	t_{\mathrm{control}},t_{\mathrm{protection}},t_{\mathrm{sync}}).
	\label{eq:event_next}
\end{equation}
Between events, topology is fixed, so a local converter can be advanced without crossing an event boundary.

For a linear or locally linear interval, the local state equation is
\begin{equation}
	\dot{x}=A_{\sigma}x+b_{\sigma}(t),
\end{equation}
and the solution over $\Delta t$ is
\begin{equation}
	x(t+\Delta t)=e^{A_{\sigma}\Delta t}x(t)
	+\int_0^{\Delta t}e^{A_{\sigma}(\Delta t-\tau)}b_{\sigma}(\tau)d\tau.
	\label{eq:event_exact_general}
\end{equation}
Thus, once the next switching instant is known, the module need not take small intermediate steps simply to reach it. Fig.~\ref{fig:pwm_events} illustrates this idea for a PWM carrier/reference crossing.

\subsection{Event-Exact Update}

This paper uses an event-exact local update. If $b_{\sigma}$ is constant over $\Delta t=t_{k+1}-t_k$,
\begin{equation}
	x_{k+1}=e^{A_{\sigma}\Delta t}x_k
	+A_{\sigma}^{-1}(e^{A_{\sigma}\Delta t}-I)b_{\sigma}.
	\label{eq:event_exact_const}
\end{equation}
The implementation may use an augmented matrix exponential to avoid forming $A_{\sigma}^{-1}$. This is practical here because each PV-module interval has fixed topology and small local dimension.

The computational distinction is
\begin{equation}
	C_{\mathrm{stepped}}\approx N_{\mathrm{steps}}C_{\mathrm{optimized\ step}},
	\label{eq:stepped_cost}
\end{equation}
where $C_{\mathrm{optimized\ step}}$ includes discretization, aggregation, reduced solves, and native execution. An event-driven method has cost
\begin{equation}
	C_{\mathrm{event}}\approx
	\sum_{m=1}^{N} N_{\mathrm{events},m}C_{\mathrm{local},m}
	+N_{\mathrm{sync}}C_{\mathrm{network}}.
	\label{eq:event_cost}
\end{equation}
The event-driven method is useful when local event advances plus occasional network synchronizations are cheaper than updating every module at every fixed step. 
\subsection{Event-Driven Execution Stack}
\begin{figure}[t]
	\centering
	\includegraphics[width=0.98\linewidth]{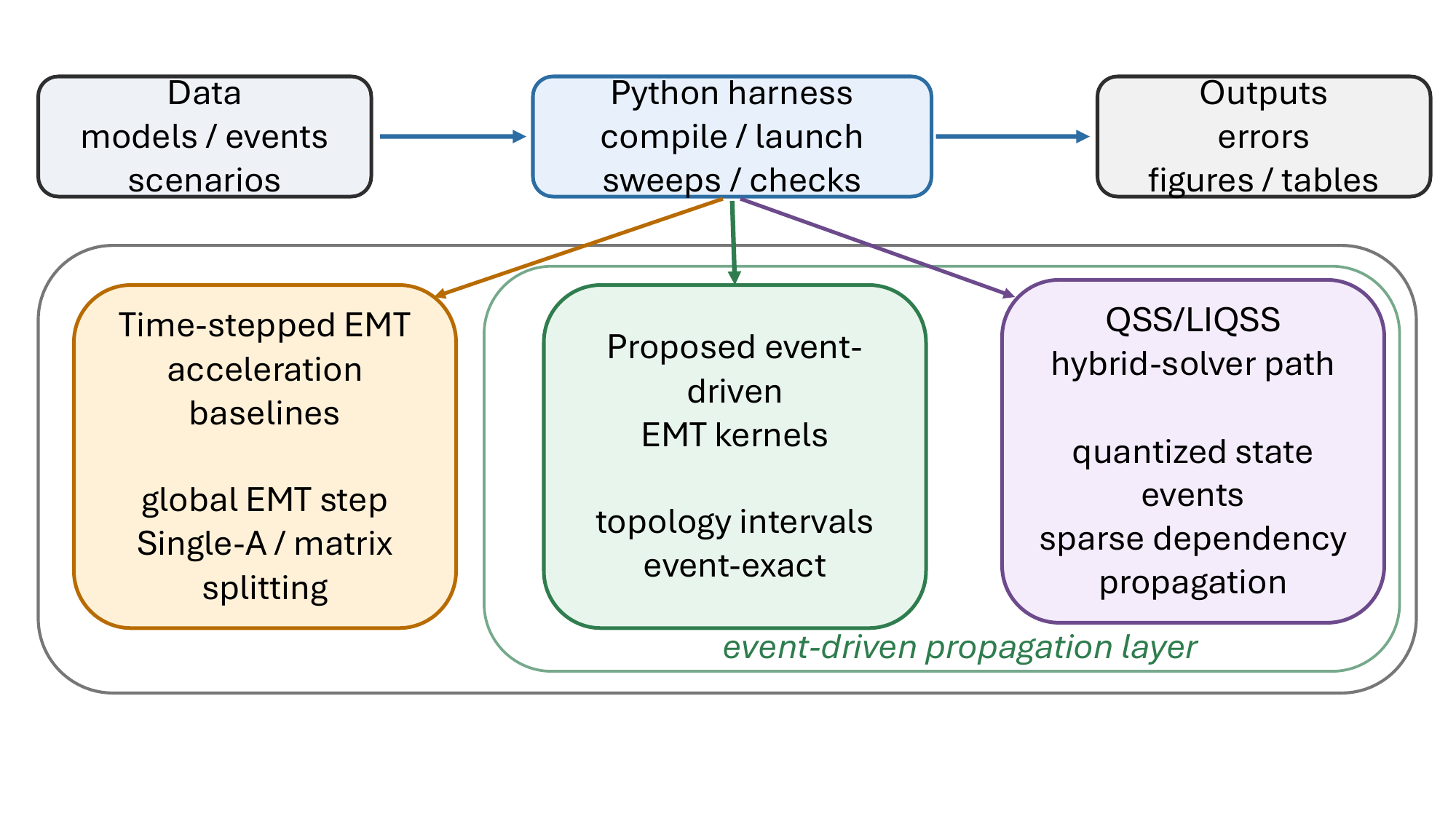}
	\vspace{-0.35cm}
	\caption{Python handles orchestration while native C kernels carry the timed event-driven simulation.}
	\label{fig:software_arch}
\end{figure}
The implementation handles deterministic PWM events directly rather than through a general priority or calendar queue. Sampled feedback controllers update the duty cycles and inverter modulation references, and those held values are used to predict the next carrier boundary or duty/reference crossing inside the current synchronization window. The module then advances from one switching/topology event to the next using the event-exact update. This gives constant work per scheduled local event and avoids queue insertion, cancellation, and rescheduling overhead for the PWM cases studied here. General event queues remain the natural extension for irregular diode crossings, protection actions, faults, and asynchronous controls \cite{nutaro}.

Fig.~\ref{fig:software_arch} summarizes the software structure. Python builds cases, compiles kernels, launches sweeps, and processes outputs; native C kernels carry the timed simulation loop. The benchmark is implemented as a distributed local/global model: each PV module owns its states and switching schedule, receives a bus voltage, and returns a current injection. During a synchronization window, bus voltages are held fixed while modules advance event to event. At the boundary, currents are accumulated at connected buses, the collector network is updated using a pre-factored solve or matrix-splitting path, and the new bus voltages are returned to the modules.

\section{Numerical Results}
\label{sec:results}
\begin{figure*}
	\centering
	\includegraphics[width=\textwidth]{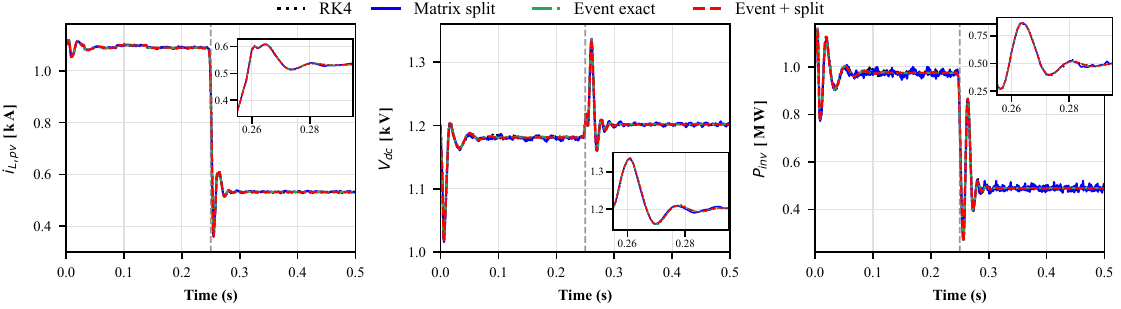}
\caption{First-module waveforms during the 1.0-to-0.5 pu power-reference step at 0.25~s.}
	\label{fig:highgran_modules}
\end{figure*}
Simulations are conducted on a MacBook Pro with 16 GB RAM and a 2.3 GHz quad-core Intel Core i7. The timed simulation loop is implemented in native C, with Python used for orchestration and post-processing. The benchmark follows the high-granularity IBR system in \cite{choi2025highgranularity}: 125 PV inverter modules behind 25 unit transformers and a multi-feeder collector network. The switched model has 1806 states: 306 collector-network states plus 125 converter modules with 12 states each. The network uses three-phase PI-line currents and bus-capacitance voltages, while each module includes PV-side, dc-link, LCL-filter, and sampled PWM feedback-control states. Choi \emph{et al.} compare a standard PSCAD Baseline with a matrix-splitting configuration, and the matrix-splitting case still takes about 0.18~h for a 0.25~s simulation \cite{choi2025highgranularity}. This motivates our approach of retaining the high-granularity matrix-splitting setting while keeping the timed simulation path in native C kernels.

We compare four native implementations: RK4, matrix splitting, event exact, and event exact + split. RK4 and matrix splitting use a 1~$\mu$s fixed local step; matrix splitting uses the collector-network update idea from \cite{choi2025highgranularity} while keeping fixed-step local propagation. Event exact advances each module over PWM-bounded topology intervals and synchronizes the collector every 10~$\mu$s. Event exact + split combines event-local module propagation with the matrix-splitting network path. Errors are normalized RMS values, $100\|y-y_{\mathrm{RK4}}\|_2/\|y_{\mathrm{RK4}}\|_2$, for the sampled first-module waveform.

Fig.~\ref{fig:highgran_modules} shows that all methods produce the same power-step transient: the event-driven traces follow RK4 closely in current, dc-link voltage, and inverter power.
\begin{table}
	\caption{Performance and error for the power-reference-step case following \cite{choi2025highgranularity}.}
	\label{tab:highgran_power}
	\centering
	\footnotesize
	\setlength{\tabcolsep}{1.5pt}
	\renewcommand{\arraystretch}{0.95}
	\begin{tabular}{@{}lrrrrrr@{}}
		\toprule
		Method & Time & Speedup & \shortstack{Local\\adv.} & \shortstack{Network\\updates} & NRMS $i_L$ & NRMS $v_{dc}$ \\
		& (s) & & (M) & (k) & (\%) & (\%) \\
		\midrule
		\shortstack[l]{Native RK4\\reference} & 41.102 & 1.0$\times$ & 62.50 & 50.0 & 0 & 0 \\
		Matrix splitting & 14.803 & 2.8$\times$ & 62.50 & 50.0 & 0.27 & 0.22 \\
		Event exact & 9.336 & 4.4$\times$ & 10.62 & 50.0 & 0.15 & 0.12 \\
		Event exact + split & 3.531 & 11.6$\times$ & 10.62 & 50.0 & 0.16 & 0.13 \\
		\bottomrule
	\end{tabular}
\end{table}
Table~\ref{tab:highgran_power} shows that matrix splitting gives 2.8$\times$ speedup, event exact gives 4.4$\times$, and event exact + split gives 11.6$\times$. Local module advances drop from 62.5 million fixed-step updates to 10.6 million event-interval advances, with current and dc-link voltage NRMS errors below 0.2\%.
\begin{figure}
	\centering
	\includegraphics[width=\linewidth]{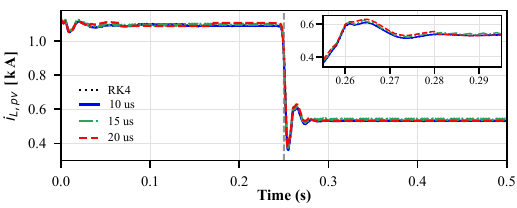}
\caption{Effect of synchronization period on first-module $i_{L,pv}$ for event exact + split.}
	\label{fig:event_sync_sweep}
\end{figure}
For event exact + split, increasing the synchronization period from 10~$\mu$s to 20~$\mu$s reduces runtime from 3.531~s to 2.684~s, improving speedup from 11.6$\times$ to 15.3$\times$, while the error increases. Fig.~\ref{fig:event_sync_sweep} compares the first module current waveform for different synchronization periods. 

\section{Conclusion}
\label{sec:conclusion}
Discrete-event methods can accelerate EMT simulation by advancing converter models only at switching events and synchronizing the network at interface boundaries. Native C kernels and event-exact updates reduce computational effort while preserving system-level behavior. Future work will integrate production sparse solvers for the network update and connect the local propagation layer to QSS/LIQSS infrastructure for asynchronous state events and sparse dependency propagation \cite{qss_review}.

\bibliographystyle{IEEEtran}
\bibliography{refs}

@ARTICLE{8086106,
  author={Li, Boyang and Zhao, Zhengming and Yang, Yi and Zhu, Yicheng and Yu, Zhujun},
  journal={CES Transactions on Electrical Machines and Systems}, 
  title={A novel simulation method for power electronics: discrete state event driven method}, 
  year={2017},
  volume={1},
  number={3},
  pages={273-282},
  doi={10.23919/TEMS.2017.8086106}}

@ARTICLE{8675318,
  author={Zhu, Yicheng and Zhao, Zhengming and Shi, Bochen and Yu, Zhujun},
  journal={IEEE Transactions on Power Electronics}, 
  title={Discrete State Event-Driven Framework With a Flexible Adaptive Algorithm for Simulation of Power Electronic Systems}, 
  year={2019},
  volume={34},
  number={12},
  pages={11692-11705},
  doi={10.1109/TPEL.2019.2907731}}

@ARTICLE{choi2025highgranularity,
  author={Choi, Jongchan and Xue, Yaosuo and Wang, Hong},
  journal={IEEE Open Access Journal of Power and Energy},
  title={Electromagnetic Transient Simulation of Large-Scale Inverter-Based Resources With High-Granularity},
  year={2025},
  volume={12},
  pages={664-677},
  doi={10.1109/OAJPE.2025.3615786}}

@book{zeigler,
author = {Bernard P. Zeigler and Alexandre Muzy and Ernesto Kofman},
title = {Theory of Modeling and Simulation, 3rd edition},
year = {2018},
publisher = {Academic Press}
}

@book{nutaro,
author = {James Nutaro},
title = {Building software for simulation},
publisher = {Wiley},
year = {2010}
}

@article{10.1177/0037549711401000,
author = {Nutaro, James and Kuruganti, Phani Teja and Protopopescu, Vladimir and Shankar, Mallikarjun},
title = {The split system approach to managing time in simulations of hybrid systems having continuous and discrete event components*},
year = {2012},
issue_date = {March     2012},
publisher = {Society for Computer Simulation International},
address = {San Diego, CA, USA},
volume = {88},
number = {3},
issn = {0037-5497},
url = {https://doi.org/10.1177/0037549711401000},
doi = {10.1177/0037549711401000},
journal = {Simulation},
month = mar,
pages = {281–298},
numpages = {18}
}

@article{qss_review,
title = {Discrete-event simulation of continuous-time systems: evolution and state of the art of quantized state system methods},
author = {Rodrigo Castro and Mariana Bergonzi and Ezequiel Pecker Marcosig and Joaquín Fernández and Ernesto Kofman},
year = {2024},
month = jun,
journal = {SIMULATION},
volume = {100},
number = {6},
pages = {613-638}
}

\end{document}